\documentclass[preprint,endfloats,prb,superscriptaddress]{revtex4}

\usepackage{amsmath,graphicx,natbib}

\newcommand{\CeCoIn}{CeCoIn$_5$}
\newcommand{\LaCoIn}{LaCoIn$_5$}
\newcommand{\LaRhIn}{LaRhIn$_5$}
\newcommand{\LaIrIn}{LaIrIn$_5$}
\newcommand{\LaMIn}{LaMIn$_5$}
\newcommand{\CeMIn}{CeMIn$_5$}

\begin{document}


\title{Crystalline Electric Field Excitations in the Heavy Fermion Superconductor CeCoIn$_5$}

\author{E. D. Bauer}
\affiliation{Los Alamos National Laboratory, Los Alamos, New Mexico 
87545, USA}

\author{A. D. Christianson}
\affiliation{Los Alamos National Laboratory, Los Alamos, New Mexico 
87545, USA}
\affiliation{University of California, Irvine, California 92697, USA}

\author{J. M. Lawrence} 
\affiliation{University of California, Irvine, California 92697, USA}

\author{E. A. Goremychkin} 
\affiliation{Argonne National Laboratory, Argonne, Illinois 60439, USA}

\author{N. O. Moreno}
\affiliation{Los Alamos National Laboratory, Los Alamos, New Mexico 
87545, USA}

\author{N. J. Curro}
\affiliation{Los Alamos National Laboratory, Los Alamos, New Mexico 
87545, USA}

\author{F. R. Trouw}
\affiliation{Los Alamos National Laboratory, Los Alamos, New Mexico 
87545, USA}

\author{J.~L. Sarrao}
\affiliation{Los Alamos National Laboratory, Los Alamos, New Mexico 
87545, USA}

\author{J. D. Thompson}
\affiliation{Los Alamos National Laboratory, Los Alamos, New Mexico 
87545, USA}


\author{R. J. McQueeney}
\affiliation{Los Alamos National Laboratory, Los Alamos, New Mexico 
87545, USA}

\author{W. Bao}
\affiliation{Los Alamos National Laboratory, Los Alamos, New Mexico 
87545, USA}

\author{R. Osborn} 
\affiliation{Argonne National Laboratory, Argonne, Illinois 60439, USA}

\date{\today}


\begin{abstract}

The crystalline electric field (CEF) energy level scheme of the heavy fermion superconductor \CeCoIn{} had been determined by means of inelastic neutron scattering (INS).  Peaks observed in the INS spectra at $\sim$ 8 meV and $\sim$27 meV with incident neutron energies between $E_i$=30-60 meV and at a temperature $T$ = 10 K correspond to  transitions from the ground state to the two excited states, respectively. The wavevector and temperature dependence of these peaks are consistent with CEF excitations.  Fits of the data to a CEF model yield the CEF parameters $B^0_2=-0.80$ meV, $B^0_4=0.059$ meV,  and $|B^4_4|= 0.137$ meV corresponding to an energy level scheme: $\Gamma_7^{(1)}$(0) [=$0.487\,|\pm 5/2> - 0.873 \,|\mp 3/2>$], $\Gamma_7^{(2)}$(8.6 meV, 100 K), and $\Gamma_6$(24.4 meV, 283 K).  
\end{abstract}


\maketitle

 
The compound \CeCoIn{} belongs to the family of CeMIn$_5$ (M=Co, Rh, Ir) heavy fermion superconductors which crystallize in the tetragonal HoCoGa$_5$ structure.\cite{Thompson03a}  The superconducting transition temperature of this material $T_c=2.3$ K is among the highest observed in a heavy fermion system.\cite{Petrovic01}  There is ample evidence for unconventional superconductivity in \CeCoIn, in which power law $T$-dependences of the specific heat and thermal conductivity are observed in the superconducting state;\cite{Movshovich01,Thompson03a} moreover, angular-dependent thermal conductivity measurements reveal a fourfold symmetry characteristic of $d$-wave superconductivity.\cite{Izawa01}   In addition, the unconventional superconductivity appears to evolve out of a non-Fermi liquid normal state deduced from the near-linear temperature dependence of the electrical resistivity $\rho$ and a logarithmic variation of the electronic specific heat coefficient $C(T)/T$, suggesting the \CeCoIn{} may be in proximity to a quantum critical point.\cite{Petrovic01,Nakatsuji02}  Knowledge of the  ground state properties is important for the understanding both of the unusual normal and superconducting states of \CeCoIn{}.  To this aim, we have carried out inelastic neutron scattering measurements on \CeCoIn{} to directly probe the crystalline electric field (CEF) splittings in this material.
 
\begin{figure}[htbp]
\includegraphics[width=4.0in]{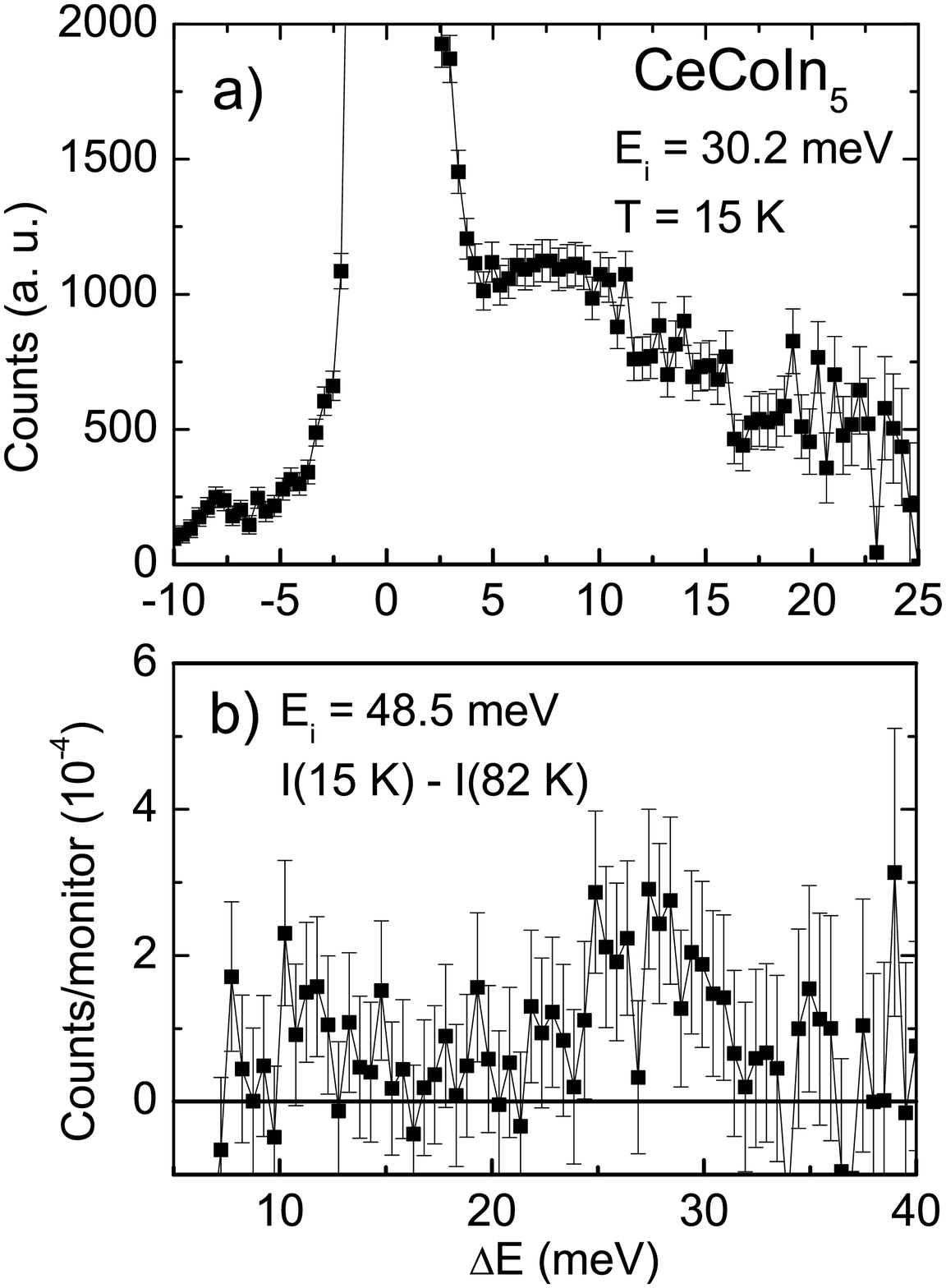}
 \caption{(a) Inelastic neutron response $I$ vs energy transfer $\Delta E$ of \CeCoIn{} at $T=15$ K and $E_i=30.2$ meV. (b) Difference spectra collected at $T=15$ K and 82 K, $I$(15 K) - $I$(82 K) vs $\Delta E$. }
\label{fig1}
\end{figure}
A polycrystalline sample ($\sim 50$ g) of \CeCoIn{} was prepared by placing stoichiometric amounts of Ce, Co, and In in an alumina crucible sealed in a quartz tube, heating to 1100 $^\circ$C, then cooling to 900 $^\circ$C at 20 $^\circ$C/hr., and finally quenching in liquid nitrogen.  The sample was then annealed at 600 $^\circ$C for 2-3 weeks.  The superconducting properties of the polycrystalline sample (i.e., $T_c=2.3$ K) were found to be consistent with single crystal results.  Various attempts to make \LaCoIn{} in a similar manner were unsuccessful. The inelastic neutron scattering measurements were performed using LRMECS at IPNS (Argonne National Laboratory) and also at PHAROS at LANSCE (Los Alamos National Laboratory).  Spectra were obtained at temperatures between $T=10-150$ K at various incident neutron energies between $E_i =30-60$ meV.  

 The inelastic neutron response of \CeCoIn{} $I$ vs energy transfer $\Delta E$ collected at PHAROS at $T=15$ K and $E_i=30.2$ meV is shown in Fig. \ref{fig1}a.  A broad hump-like feature is visible in the data at $\Delta E \sim 8$ meV suggesting a crystalline electric field excitation at that energy.  Data collected using a somewhat higher incident energy $E_i=48.5$ meV at $T=10$ K and $82$ K, plotted as the difference of the spectra at the two temperatures $I$(15 K) - $I$(82 K) vs $\Delta E$ (Fig. \ref{fig1}b), reveal another feature centered at $\sim$27 meV (the peak at 8 meV is not visible in this spectra due to the poorer energy resolution at $E_i=48.5$ meV).  The temperature dependence of the peak at $\sim$ 27 meV, i.e., decreasing intensity with increasing $T$, suggests that it corresponds to a CEF transition from the ground state to an  excited state, and not to a phonon excitation.  The same conclusion can be drawn from the $T$-dependence of the 8 meV peak in data collected at LRMECS at $E_i=35$ meV (not shown).  Further evidence for  CEF transitions from the ground state to excited states at $\sim 8$ meV and $\sim27$ meV in \CeCoIn{} is provided by the wavevector dependence of these features, which  corresponds to that of the Ce$^{3+}$ form factor.  
\begin{figure}[htbp]
\includegraphics[width=5.0in]{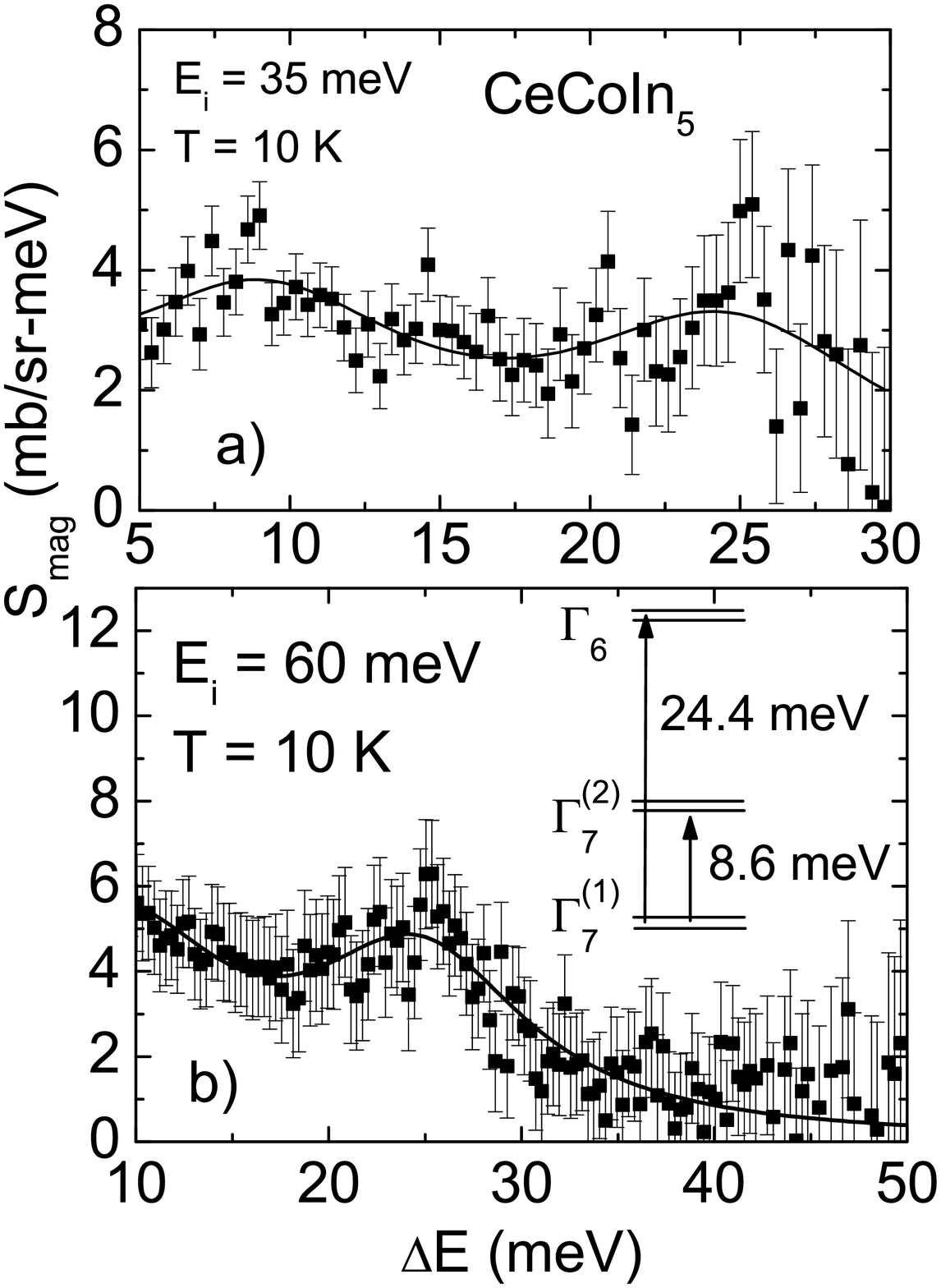}  
 \caption{Magnetic contribution to the inelastic neutron scattering of \CeCoIn{} at $T=10$ K for (a) $E_i$=35 meV and (b) $E_i$=60 meV.  The solid lines are fits of the data to a CEF model yielding the CEF energy level scheme in (b).}
\label{ceffits}
\end{figure}

The CEF Hamiltonian in tetragonal symmetry is given by:  \(\mathcal{H}_{CEF}=B^0_2 \, O^0_2 + B^0_4 \, O^0_4 + B^4_4 \, O^4_4,\) where $B^m_l$  and  $O^m_l$ are the CEF parameters and Stevens' operators, respectively. The six-fold degenerate Ce$^{3+}$ $J=5/2$ multiplet splits into 3 doublets:  $\Gamma_7^{(1)}=\alpha \,|\pm 5/2> - \beta \,|\mp 3/2>$, $\Gamma_7^{(2)}=\beta \,|\pm 5/2> + \alpha \,|\mp 3/2>$, and $\Gamma_6=\,|\pm 1/2>$ under the influence of the crystalline electric field.\cite{Fischer87a}

The magnetic contribution to the inelastic neutron scattering response of \CeCoIn{}  at $T=10$ K is shown in Fig. \ref{ceffits}a for $E_i=35$ meV and Fig. \ref{ceffits}b for $E_i=60$ meV.  The nonmagnetic contribution inferred from measurements on  \LaRhIn{} ($E_i=35$ meV, Fig. \ref{ceffits}a) or \LaIrIn{} ($E_i=60$ meV, Fig. \ref{ceffits}b) has been subtracted from the low-angle scattering data of \CeCoIn{}  using the expression \(S_{mag} = S(\textnormal{Ce},20 ^\circ) -  f S(\textnormal{La},20 ^\circ)\), where $f$ is the ratio (0.75 for Rh, 0.59 for Ir) of the total scattering cross section $\sigma (\textnormal{\CeCoIn})/\sigma (\textnormal{\LaMIn})$ (other background subtractions, i.e., YCoIn$_5$, yield similar results\cite{Christianson03a}).  The data have also been normalized to a vanadium standard to put the scattering on an absolute scale.  A simultaneous least-squares fit of the two datasets to a CEF model yields the CEF parameters for \CeCoIn: $B^0_2=-0.80(7)$ meV, $B^0_4=0.059(3)$  meV,  and $|B^4_4|= 0.137(1)$ meV (Fig. \ref{ceffits}).  Additional fit parameters include:  1) an overall scale factor ($F=0.73$ for $E_i=35$ meV, $F=0.90$ for $E_i=60$ meV), a Lorentzian  width $\Gamma$=6.5 meV of the CEF excitations, and 3) a quasielastic width $\Gamma_{QE}=1.6$ meV fixed to be (1/4)$\Gamma$ in accord with the value determined from NMR measurements of the spin-lattice relaxation rate.\cite{Curro03}  (The fit also includes the effects of instrumental resolution.)   The resulting CEF level scheme from this fit for \CeCoIn{} is:   $\Gamma_7^{(1)}$[0], $\Gamma_7^{(2)}$[8.6(5) meV, 100 K], and $\Gamma_6$[24.4(7) meV, 283 K] (Fig. \ref{ceffits}b), with a mixing parameter $\beta=0.87(5)$.  The magnetic susceptibility is reasonably well-described by this CEF level scheme above $\sim75$ K including an isotropic molecular field constant ($\lambda= 50$ mol/emu) to account for the Kondo effect and magnetic correlations in this material (not shown).  The systematic errors due to the large background contribution and significant absorption are estimated to be larger than the statistical uncertainties of the various parameters listed above;  a more complete analysis similar to that reported in Ref. \onlinecite{Christianson02a} on the \CeCoIn{} system will be presented elsewhere.\cite{Christianson03a}


Three other reports of the CEF properties of \CeCoIn{} exist in the literature.
Nakatsuji  {\it{et al.}}\cite{Nakatsuji02} determined the following scenario based on $\chi(T)$ and $C(T)$ measurements: $\Gamma_7^{(1)}$(0), $\Gamma_7^{(2)}$(148 K), and $\Gamma_6$(197 K) with  $\beta=$0.92.   Another CEF scheme determined from $\chi(T)$ by Shishido {\it{et al.}}\cite{Shishido02} has been reported:  $\Gamma_7^{(1)}$(0), $\Gamma_7^{(2)}$(151 K), and $\Gamma_6$(166 K) with  $\beta=$0.52. A CEF analysis of $\chi(T)$ by Pagliuso {\it{et al.}}\cite{Pagliuso02} yield a different CEF level scheme: $\Gamma_6$(0 K), $\Gamma_7^{(2)}$(35 K), and $\Gamma_7^{(1)}$(72 K), with  $\beta$=0.1. [The labels of Ref. \onlinecite{Pagliuso02}, $\Gamma_7^{(1)}$ and $\Gamma_7^{(2)}$, have been switched to compare with the convention used in this paper].  
The relative level scheme and degree of mixing (i.e., $\beta=0.87$) that we observe are  more consistent with that of  Nakatsuji  {\it{et al.}}\cite{Nakatsuji02} and Shishido {\it{et al.}}\cite{Shishido02}. In addition, a negative value of $B^0_2$, consistent with the anisotropy of $\chi(T)$ of \CeCoIn{} discussed below, indicates a  $\Gamma_7^{(1)}$ or $\Gamma_7^{(2)}$ ground state rather than the $\Gamma_6$ ground state scenario proposed in Ref. \onlinecite{Pagliuso02}.

The family of \CeMIn{} heavy fermion superconductors, in which the tetragonal structure is comprised of alternating layers of CeIn$_3$ separated by a layer of MIn$_2$ along the $c$-axis, can be considered to be the two-dimensional (2D) analogue of another heavy fermion superconductor CeIn$_3$.  As a result of the lower tetragonal symmetry in the ``115" structure, the excited quartet of the doublet-quartet ($\Gamma_7$-$\Gamma_8$) crystalline electric field scheme of CeIn$_3$ splits into two doublets.  The effects of the CEF in this tetragonal system contributes, at least in part, to anisotropy in the magnetic susceptibility.  A simple estimate of the $B^0_2$ CEF parameter is provided by the anisotropic Curie-Weiss temperatures (assuming no magnetic exchange) and is given by: \( \theta_{ab} - \theta_c = [3(2J-1)(2J+3)/10]B^0_2\).\cite{Boutron73}  Using $\theta_{ab}=-80$ K and $\theta_c=-56$ K, determined from fits of $\chi(T)$ to a Curie-Weiss law above $\sim200$ K, a value $B^0_2=-0.22$ meV is obtained.  This is in reasonable agreement with the value determined from the CEF analysis of the INS data ($B^0_2=-0.80$ meV) considering the strong Kondo interactions present in \CeCoIn.  It has been postulated that the order of magnitude increase in $T_c$ from CeIn$_3$ ($T_c \sim 0.2$ K) to \CeCoIn{}  ($T_c=2.3$ K) is related to the quasi-2D nature of the \CeMIn{} compounds, in which the increase in anisotropy increases the 2D character of the spin fluctuations relevant for superconductivity;\cite{Pagliuso02}  this idea is supported by theoretical work by Monthoux and Lonzarich.\cite{Monthoux99}  It is less clear to what extent the crystalline electric fields influence the superconductivity in the \CeMIn{} materials.  Recent theoretical work by Hotta and Ueda\cite{Hotta03} suggest that the shape of the $f$-orbitals under the influence of the CEF may affect the orbital fluctuations that contribute to superconductivity.  Further experimental and theoretical efforts are needed to address this issue.

Work at Los Alamos was performed under the auspices of the DOE. The work has also benefited from the use of the Los Alamos Neutron Science Center at Los Alamos National Laboratory.  LANSCE is funded by US Department of Energy under Contract W-7405-ENG-36.
     


\end{document}